\documentclass[12pt]{article}

\usepackage{graphicx}
\usepackage{amsmath, amssymb}
\usepackage{physics}
\usepackage{cite}
\usepackage{hyperref}
\usepackage{geometry}
\usepackage{caption}
\usepackage[format=plain, labelfont=bf]{caption}
\usepackage{placeins}
\usepackage{float}
\usepackage[protrusion=true,expansion=true]{microtype}

\title{Machine learning assisted High-Throughput study of M$_4$X$_3$T$_x$ MXenes}
\author{
Sakshi Goel$^{1}$ and Arti Kashyap$^{1,*}$ \\
\small $^{1}$School of Physical Sciences, Indian Institute of Technology Mandi, 
Himachal Pradesh, India 175075 \\
\small $^{*}$Email: arti@iitmandi.ac.in
}
\date{}

\begin{document}

\maketitle

\begin{abstract}
\normalsize
In this work, we employ a machine-learning-assisted high-throughput density functional theory framework to systematically investigate the stability, electronic structure, and magnetic ground states of 234 M$_4$X$_3$T$_x$ MXenes. The machine learning model predicts lattice parameters with up to 94\% accuracy using a relatively small training dataset and significantly reduces structural optimization time in high-throughput calculations. Based on total energy and density-of-states analyses, we classify the magnetic nature of MXenes across different transition-metal compositions and surface terminations. Ti-, Zr-, Hf-, Nb-, and Ta-based MXenes are found to be non-magnetic metals for all functional groups considered, while Sc- and Y-based systems exhibit a range of behaviors including weak ferromagnetism and semiconducting character. V- and Fe-based MXenes are identified as antiferromagnetic metals, whereas Cr- and Mn-based MXenes yield 16 ferromagnetic systems with spin polarization ranging from 50\% to 100\%. 
\end{abstract}
\vspace{1em} 

\section{Introduction}

Since the experimental realization of the two-dimensional titanium carbide Ti$_3$C$_2$T$_x$ in 2011, MXenes have emerged as one of the most rapidly expanding families of two-dimensional materials, exhibiting exceptional potential for applications in energy storage, MXetronics and spintronics due to their unique layered microstructure and the presence of transition metals \cite{Amrillah2023, Li2022, Kim2020}. MXenes are described by the general formula M$_{n+1}$X$_n$T$_x$ ($n = 1$--$4$), where M is an early transition metal, X is carbon or nitrogen, and T$_x$ denotes surface functional groups. They are typically synthesized by selectively etching the A layer from their parent MAX phases, M$_{n+1}$AX$_n$ where A is typically a group 13 or 14 element \cite{Naguib2021}. Their layered structure, combined with the presence of transition metals and chemically active surfaces, gives rise to a remarkable degree of compositional and structural tunability.

Over the past decade, this tunability has been significantly expanded through both experimental and theoretical advances, including the synthesis of bimetallic MXenes \cite{He2019}, the experimental realization of higher-order MXenes such as the M$_4$X$_3$T$_x$  and M$_5$X$_4$T$_x$ families \cite{Hussain2024}, the identification of previously unexplored and mixed surface terminations, and the stabilization of novel stoichiometries~\cite{Wahib2026}. These developments have substantially enlarged the MXene design space, leading to the expectation that the number of thermodynamically stable MXenes will grow rapidly with continued progress. However, the resulting combinatorial explosion of possible elemental compositions, layer thicknesses, and surface chemistries far exceeds what can be explored using conventional trial-and-error strategies or isolated first-principles calculations~\cite{Gouveia2025}.

In this context, the integration of automated high-throughput density functional theory (HT-DFT) with data-driven and machine-learning (ML) approaches has emerged as a powerful framework for accelerating MXene discovery and property prediction \cite{Kashyap2024, Xin2023, Rittiruam2022}. A major computational bottleneck in HT-DFT studies of unexplored materials is the optimization of lattice parameters, which can vary substantially with elemental substitution even within the same material family. Since the accuracy of lattice constants critically influences predicted electronic, magnetic, and transport properties, inefficient structural optimization severely limits large-scale screening \cite{Shruthi2026, Zhang2020}. Previous studies have demonstrated that ML models can predict lattice parameters with high accuracy, often exceeding 90--95\%, thereby providing reliable initial structures and significantly reducing computational cost~\cite{Jiang2024, Su2023, Jarin2022, Li2021, Miyazaki2021, Zhang2020}.

Despite the rapid expansion of MXene research, magnetism in MXenes remains comparatively underexplored. While extensive studies have focused on the magnetic properties of $n = 1$ and $n = 2$ MXenes~\cite{Scheibe2019, He2019, Yang2018, Kumar2017, Li2017, Wang2016}, higher-order MXenes ($n = 3$ and $4$) have received limited attention, primarily due to synthesis challenges. Nevertheless, recent experimental reports have confirmed the successful synthesis of several $n = 3$ MXenes, including Ti$_4$N$_3$, Nb$_4$C$_3$, Mo$_2$Ti$_2$C$_3$, Ta$_4$C$_3$, V$_4$C$_3$T$_x$, and V$_3$CrC$_3$~\cite{Hussain2024, Urbankowski2016, Ghidiu2014 }, highlighting the feasibility and relevance of this class. Importantly, recent ML-driven studies have predicted several $n = 3$ MXenes, such as Mn$_4$N$_3$Cl$_2$, Cr$_4$N$_3$Cl$_2$, and Mn$_4$N$_3$S$_2$, to be both thermodynamically stable and magnetic, suggesting that higher-order MXenes may host robust magnetism and high spin polarization~\cite{Khatri2023, Xin2023}.

Motivated by these developments, we employ a combined ML-assisted high-throughput DFT (HT-DFT) framework to systematically investigate the structural stability, electronic structure, and magnetic properties of 234 M$_4$X$_3$ and M$_4$X$_3$T$_x$ MXenes, where M = (Fe, Sc, Mn, Ti, V, Cr, Y, Zr, Nb, Mo, Hf, Ta), X = C or N, and T$_x$ = (O, F, Cl, Br, I, S, Se, Te). ML-predicted lattice parameters are incorporated as initial inputs within the HT-DFT workflow, leading to a substantial reduction in computational cost for structural relaxation without compromising accuracy. For functionalized MXenes, multiple adsorption geometries are examined to identify the most stable surface configuration, while one ferromagnetic (FM) and three antiferromagnetic (AFM) arrangements are considered to determine the magnetic ground state. Our results reveal several stable $n = 3$ MXenes with high spin polarization, including multiple half-metallic candidates, highlighting the strong potential of higher-order MXenes for spintronic applications. Although the present dataset is moderate in size, the proposed ML-assisted HT-DFT framework provides a basis for future extension to larger and unexplored MXene chemical spaces.

\FloatBarrier
\begin{figure}[H]
    \centering
    \includegraphics[width=1\linewidth]{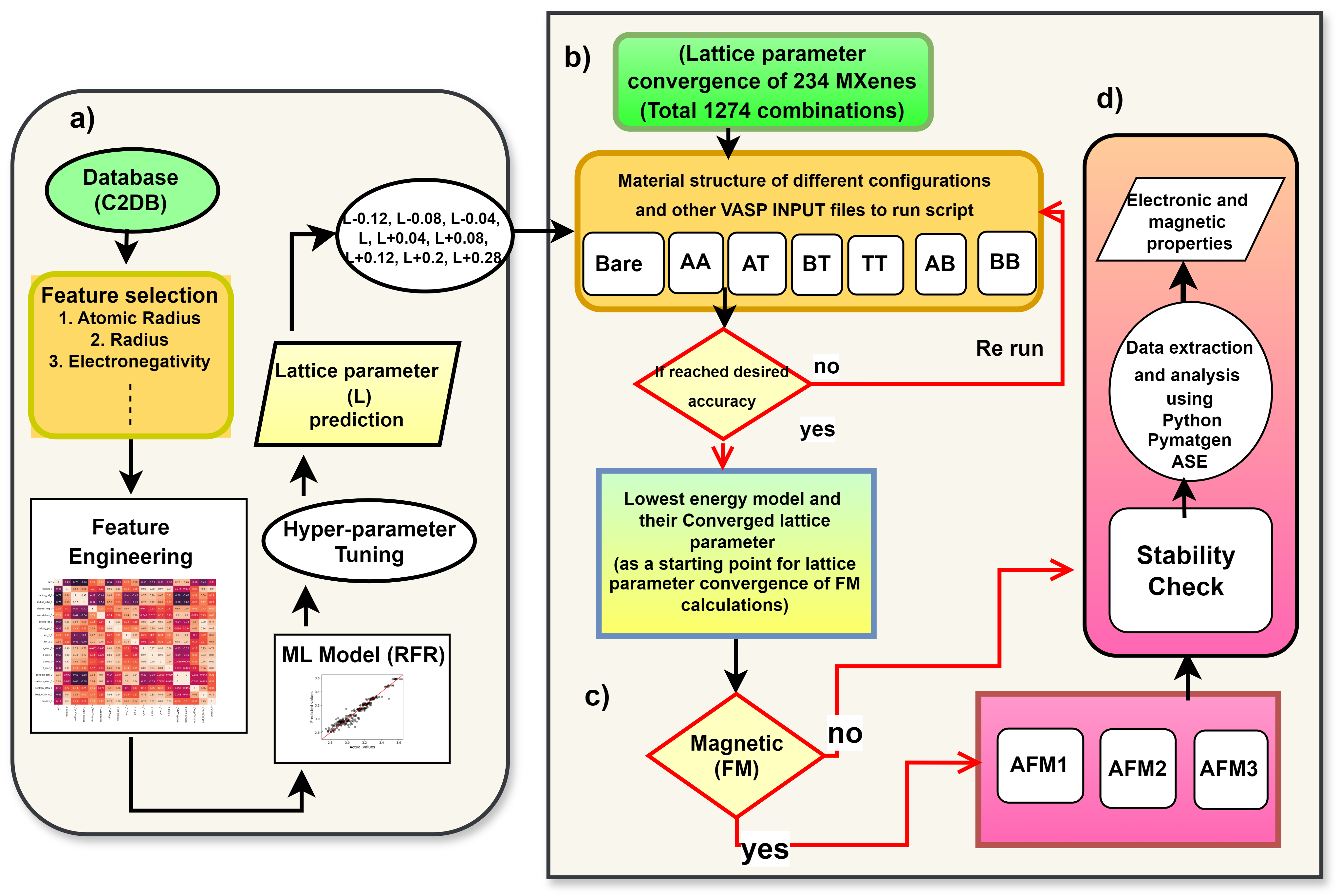}
    \caption{An overview of the workflow: (a) Lattice constant (L) prediction of 234 MXenes using ML, (b) lattice constant optimization for M$_4$X$_3$ and six M$_4$X$_3$T$_x$ configurations using ML predicted L as a starting point, (c) magnetic ground state calculation, (d) data analysis.}
    \label{fig:ml}
\end{figure}

\section{Methodology}
\subsection{Machine Learning }

The schematic representation of the methodology is illustrated in Figure~\ref {fig:ml}. We have used the Computational 2D Materials Database (C2DB) to train the machine learning model \cite{Gjerding2021}. A total of 275 materials were classified as MXenes in the C2DB database. The 275 MXenes from C2DB contain a combination of different types of MXenes; therefore, it can be applied to predict cell parameters of M$_4$X$_3$T$_x$ MXenes and other combinations. We initially selected 17 elemental features, namely atomic radius, Van Der Waals radius, atomic mass, first and second ionization energy, electronegativity, number of valence electrons, heat of formation, density, electron affinity, melting point, boiling point, atomic number and number of electrons in different orbitals (s,p,d,f). However, 6 features (boiling point, atomic number and number of electrons in different orbitals (s,p,d,f)) were found to have a Pearson correlation coefficient of more than 0.95 during correlation analysis and were dropped. Therefore, the remaining 11 are used as input features. Pymatgen library \cite{Ong2013} was used to collect the data of these 275 MXenes materials from C2DB. The elemental features were extracted using the Mendeleev Python library \cite{mendeleev2014}. Sci-kit learn library is used to apply various ML algorithms \cite{Pedregosa2011}. The grid search cross-validation method is used for hyperparameter tuning of ML algorithms. We also assessed the ML model’s performance using 5-fold cross-validation.

\FloatBarrier
\begin{figure}[H]
    \centering
    \includegraphics[width=0.8\linewidth]{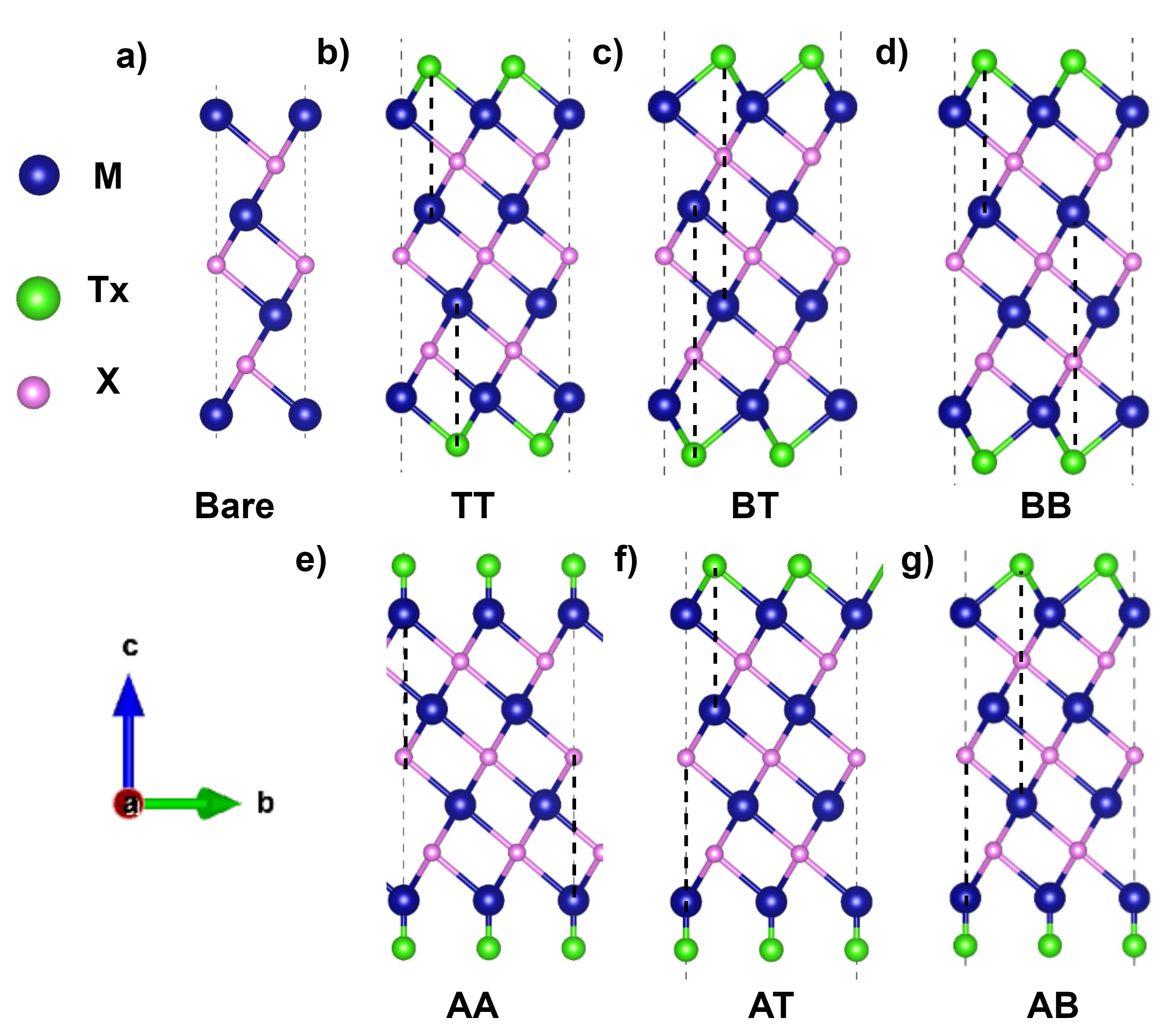}
    \caption{Geometrical structures of the bare and functionalized MXenes (at different positions): Side view of (a) Bare MXene, (b--d) functional group attached at the top and bottom site of M, (e) at the site of X, and (f--g) at M and X sites}
    \label{fig:model}
\end{figure}

\subsection{Density functional theory}

Figure.~\ref{fig:model}(a) presents the geometrical structure of the bare M$_4$X$_3$ MXene, and Figure.~\ref{fig:model}(b-g) presents the functionalized MXenes, constructed by considering the hollow sites (TT, BT, BB) and hollow and top sites (AA, AT, AB) of Cr and C.
The density functional theory (DFT) calculations of these structures were performed using Vienna ab initio simulation package (VASP) \cite{Kresse1996},\cite{Kresse1996PW}. Projector augmented wave (PAW) \cite{Kresse1996UP} pseudopotentials with a cutoff energy of 620 eV were used to describe the electron-electron interaction. The generalized gradient approximation (GGA) with the Perdew Burke-Ernzerhof (PBE) was used to treat the exchange-correlation effects \cite{Perdew1996}. The convergence criteria for energy were set to 10$^{-6}$~eV/atom and the atomic positions were optimized until the forces acting on each atom were reduced to 0.01 eV/~\AA. A $12 \times 12 \times 1$ Gamma-centered k-point mesh was considered for geometry optimization. The on-site Coulomb interactions were introduced using Dudarev’s GGA+U approach \cite{Dudarev1998} to describe the strong-correlation effects in transition-metal atoms. The Hubbard U values for different transition metal atoms were selected from previous studies and are 4eV for Cr, Fe, Mn, and 3eV for V \cite{Kumar2017}, \cite{Wang2017}. To prevent any interactions between the adjacent layers, a vacuum of 2~nm is considered. The Pymatgen and Atomic Simulation Environment (ASE) \cite{Hjorth2017} libraries are utilized to plot the DOS and extract structural information of the materials.

\section{Results and Discussion}
\subsection{ML predicted and DFT calculated Lattice parameter Comparison}
In this section, we compared the lattice parameters calculated using ML and DFT (with ML-predicted values as the starting guess). For the selected 275 materials classified as MXenes from the (C2DB) database \cite{Gjerding2021}, machine learning models are trained and compared. To train the data, two ML algorithms, Random Forest Regressor (RFR) and XGBoost (XGB) are applied in this study. RFR has been found to perform better in previous study for cell parameter prediction \cite{Su2023} and XGB is generally considered more accurate than RFR due to its ability to learn from previous errors. The hyperparameter tuning is performed using GridSearchCV and the tuned parameters are: for RFR (n\_estimators = 80, min\_samples\_leaf = 4, max\_features =\texttt{auto}, max\_depth = 7, bootstrap = True) and XGB (colsample\_bytree = 0.7980, gamma = 0.001, learning\_rate = 0.1688, max\_depth = 5, n\_estimators = 140, subsample = 0.862). The performance of machine learning models is examined by R$^2$ score, Mean Absolute Error (MAE), and Mean Absolute Percentage Error (MAPE) metrics. The R$^2$ Score, MAE and MAPE for RFR are 0.89, 0.04 and 1.29\% and for XGB 0.87, 0.042, and 1.52\%, respectively. The results show that RFR accuracy is higher than XGB. Therefore, the lattice constant of 234 compositions is predicted using RFR. 
\\
In the next step, ML-predicted lattice parameters were used as the initial guess in the HT-DFT pipeline to achieve convergence. The minimum and maximum deviations between ML-predicted and HT-DFT calculated lattice constants (for the lowest-energy model of functionalized MXenes) are 0 and 0.28~\AA, respectively. Mo- and W-based nitride MXenes did not converge properly and were therefore excluded from the study. Out of 217 compounds, 64 exhibit a difference of 0--0.05~\AA, 62 have 0.05--0.10~\AA, 45 show 0.10--0.15~\AA, and 46 compounds display a larger difference of 0.15--0.28~\AA. These results indicate that most compounds have a lattice mismatch within 0--6\%. 
Compounds with a lattice mismatch of 6--9\% generally correspond to systems for which GGA+$U$ calculations were performed. It is well known that GGA+$U$ tends to overestimate the cell parameters by 2--3\% compared with the GGA approximation~\cite{Lutfalla2011, Chibani2022}. Neglecting this overestimation, the accuracy of the RFR model for lattice constants is approximately 94\%. The use of ML-predicted cell parameters as the initial guess dramatically reduced the computational cost of structural convergence, demonstrating the effectiveness of integrating machine learning with HT-DFT workflows for large-scale materials screening. The results demonstrate that the HT-DFT calculated lattice constants closely match those reported in earlier studies(see Table~\ref{tab:lattice_constants}).

\begin{table}[htbp]
\centering
\caption{A comparison of ML predicted, HT-DFT calculated and previously reported lattice parameter (units in \AA)}
\label{tab:lattice_constants}
\begin{tabular}{lccc}
\hline
Composition & ML-predicted & HT-DFT calculated & Previously reported \\
\hline
Sc$_4$C$_3$      & 3.33  & 3.410 & 3.411~\cite{Chibani2022} \\
Ti$_4$C$_3$      & 3.058 & 3.098 & 3.093~\cite{Chibani2022} \\
Zr$_4$C$_3$      & 3.327 & 3.327 & 3.356~\cite{Chibani2022} \\
Mo$_4$C$_3$      & 3.112 & 3.072 & 3.077~\cite{Chibani2022} \\
Hf$_4$C$_3$      & 3.305 & 3.285 & 3.296~\cite{Chibani2022} \\
W$_4$C$_3$       & 3.136 & 3.086 & 3.086~\cite{Chibani2022} \\
Y$_4$C$_3$       & 3.581 & 3.621 & 3.83~\cite{Aliakbari2022} \\
Nb$_4$C$_3$      & 3.142 & 3.142 & 3.14~\cite{Huang2023} \\
Ti$_4$N$_3$      & 3.056 & 2.976 & 2.993~\cite{Urbankowski2016} \\
V$_4$C$_3$       & 2.947 & 2.947 & 2.92~\cite{Peng2022,Huang2023} \\
Ti$_4$C$_3$F$_2$ & 3.059 & 3.059 & 3.084~\cite{Xie2013} \\
Ti$_4$N$_3$F$_2$ & 3.046 & 3.046 & 3.027~\cite{Xie2013} \\
\hline
\end{tabular}
\end{table}

\subsection{Lowest Energy Configurations for functionalized MXenes}

Considering that functional groups can be attached at different positions on bare MXenes, six distinct configurations were examined in this study, labeled TT, BT, BB, AA, AT, and AB, as shown in Figure.~\ref{fig:model}(b-g). Geometry optimization was performed separately for all $\sim$1300 (210$\times$6) configurations. The results indicate that TT and BT are the most prevalent configurations in M$_4$X$_3$T$_x$ MXenes. Among 192 functionalized MXenes, 123 adopt the TT configuration, while 56 are in the BT configuration. Only 13 compounds were found in the AA and AT configurations: W$_4$C$_3$Br$_2$ (AT), Fe$_4$C$_3$I$_2$, V$_4$C$_3$I$_2$, Cr$_4$C$_3$I$_2$, Cr$_4$C$_3$Br$_2$, Cr$_4$C$_3$Te$_2$, Mo$_4$C$_3$F$_2$, Mo$_4$C$_3$I$_2$, W$_4$C$_3$F$_2$, W$_4$C$_3$Cl$_2$, Fe$_4$N$_3$I$_2$, Fe$_4$N$_3$Te$_2$, and Mn$_4$N$_3$I$_2$. For these I- and Te-functionalized MXenes, the TT, BT, and AA configurations exhibit small energy differences of approximately 10--20~meV. Thus, apart from the AA configuration, these MXenes could also be stable in the BT and TT configurations and may display distinct properties. In our study, the TT model exhibits the lowest energy for F- and O-functionalized Ti-based MXenes, which is consistent with previous reports~\cite{Xie2013, Urbankowski2016}. Also, note that further studies are performed only for the lowest energy configuration.

\subsection{Magnetic ground state study}

\begin{figure}[H]
    \centering
    \includegraphics[width=1\linewidth]{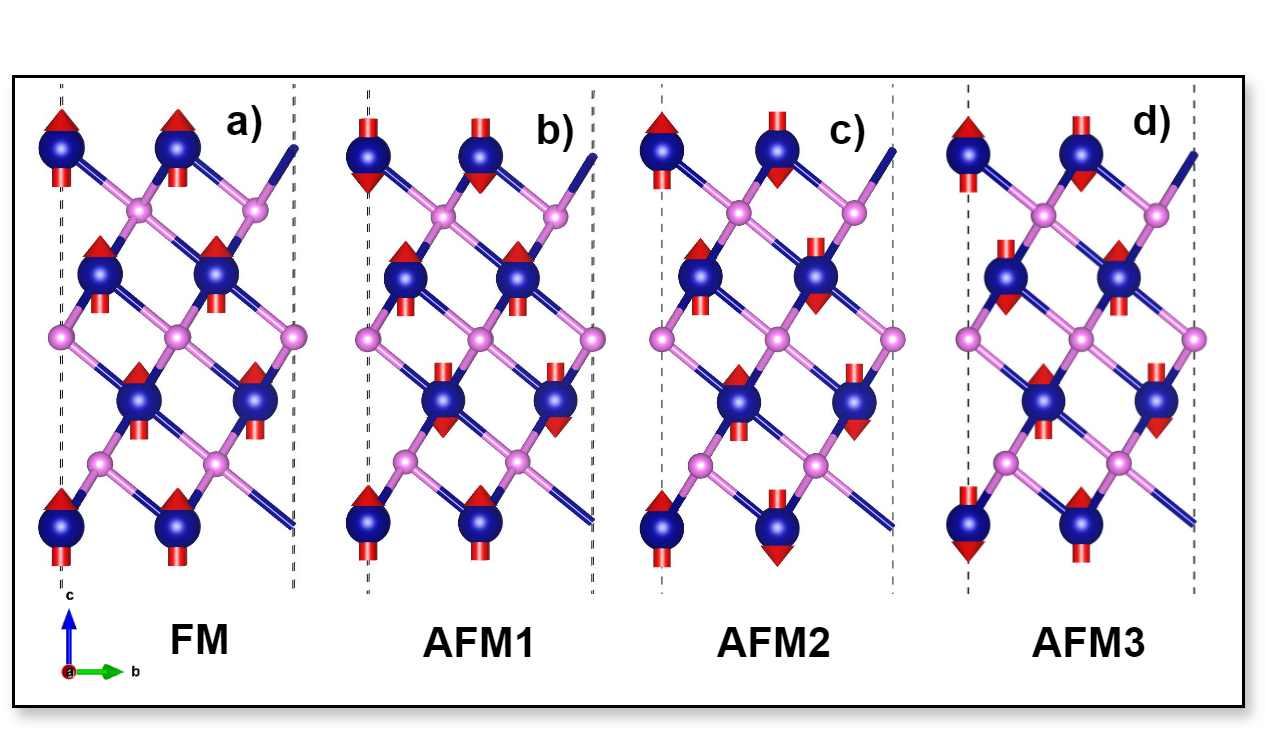}
    \caption{Side view of different magnetic configurations. (a) FM, (b) AFM1, (c) AFM2, (d) AFM3, the Up arrow represents spin up and down to spin down.}
    \label{fig:mag1}
\end{figure}

After investigating the convergence and most stable configuration of functionalized MXenes, we understand the magnetic ground state of the materials. For magnetic calculations, one FM and three AFM configurations (AFM1, AFM2, AFM3), as shown in Figure~\ref{fig:mag1}, are considered. For AFM3 and AFM2 structures, calculations are performed on a 2×2×1 supercell. The results of bare MXenes show that Sc$_4$C$_3$, Mo$_4$C$_3$, W$_4$C$_3$, Nb$_4$N$_3$, and Ta$_4$N$_3$ are nonmagnetic, while the remaining 19 bare MXenes exhibit magnetic behavior. Among the magnetic bare MXenes, only five compounds—Cr$_4$C$_3$, V$_4$C$_3$, Y$_4$C$_3$, Zr$_4$C$_3$, and Hf$_4$C$_3$ possess a ferromagnetic ground state, with total magnetic moments of 11.8, 4.9, 0.135, 0.72, and 0.92~(units in $\mu_\mathrm{B}$ per unit cell), respectively, while others are antiferromagnetic.

\begin{figure}[H]
    \centering
    \includegraphics[width=0.5\linewidth]{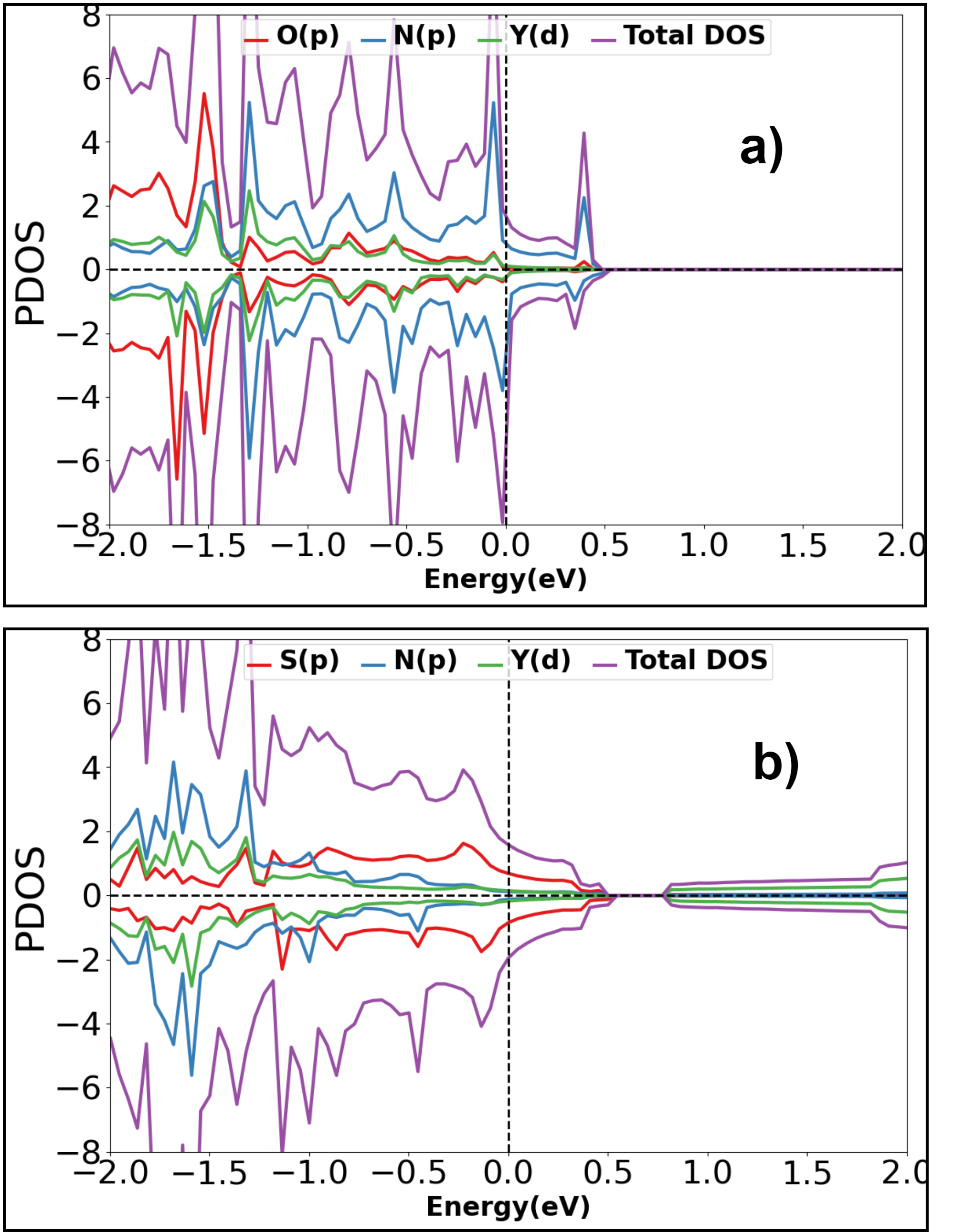}
    \caption{ Projected Density of States for weak ferromagnetic materials (a) Y$_4$N$_3$O$_2$ (b) Y$_4$N$_3$S$_2$}
    \label{fig:magnetic}
\end{figure}

However, upon functionalization with F, Cl, Br, I, O, S, Se, and Te, we find that Sc-, Ti-, Y-, Zr-, Nb-, Hf-, and Ta-based MXenes are non-magnetic for all surface terminations, except for Y$_4$C$_3$O$_2$, Y$_4$N$_3$O$_2$, and Y$_4$N$_3$S$_2$. These MXenes exhibit a FM ground state with small total magnetic moments of 0.5~$\mu_\mathrm{B}$, 0.052~$\mu_\mathrm{B}$, and 0.129~$\mu_\mathrm{B}$, per unit cell, respectively. Among them, Y$_4$C$_3$O$_2$ shows the largest magnetic moment. Local magnetic moments reveal that the dominant contribution to the magnetic moment arises from the C~$p$ orbitals in Y$_4$C$_3$O$_2$, whereas in Y$_4$N$_3$O$_2$ and Y$_4$N$_3$S$_2$, the main contribution originates from the $p$ orbitals of N and S, respectively. This behavior can be understood by considering that after donating the unpaired electrons to nearby C atoms, no unpaired electron is left on the Y d-orbital in the Y$_4$C$_3$ case. Here, the oxidation state of Y, C and N is considered as Y$^{3+}$, C$^{4-}$ and N$^{3-}$. Upon oxygen passivation, the higher electronegativity of O compared to C results in enhanced charge withdrawal, increasing the spin polarization on C atoms and thereby leading to a higher magnetic moment in Y$_4$C$_3$O$_2$. In contrast, in Y$_4$N$_3$-based MXenes, residual $d$ electrons on Y are shared with the O and S functional groups on surface passivation. Consistent with the electronegativity trend O $>$ N $>$ S, the N-$p$ orbitals contribute more to the magnetic moment in Y$_4$N$_3$O$_2$, while the S-$p$ orbitals dominate in Y$_4$N$_3$S$_2$.This interpretation is further supported by the strong Y-$d$–O-$p$ hybridization near the Fermi level in Y$_4$N$_3$O$_2$, as well as Y-$d$–N-$p$ hybridization in Y$_4$N$_3$S$_2$, as shown in Figure.~\ref{fig:magnetic}(a,b). The energy differences between the non-magnetic and ferromagnetic states are 3~meV, 1~meV, and 16~meV for Y$_4$N$_3$O$_2$, Y$_4$N$_3$S$_2$, and Y$_4$C$_3$O$_2$, respectively, confirming the weak ferromagnetic nature of these systems. Interestingly, similar magnetic behavior is not observed in Sc-based MXenes or in Se- and Te-terminated Y-based MXenes, despite Sc and Y belonging to the same periodic group.

Further, in the search for ferromagnetic (FM) materials, we identify that Mn$_4$C$_3$F$_2$, Mn$_4$C$_3$Cl$_2$, Mn$_4$C$_3$I$_2$, Mn$_4$C$_3$Br$_2$, Mn$_4$C$_3$S$_2$, Mn$_4$C$_3$Se$_2$, Mn$_4$C$_3$Te$_2$, Cr$_4$C$_3$O$_2$, Cr$_4$C$_3$F$_2$, Cr$_4$C$_3$Cl$_2$, Cr$_4$C$_3$S$_2$, Cr$_4$C$_3$Se$_2$, Mn$_4$N$_3$O$_2$, Mn$_4$N$_3$F$_2$, Mn$_4$N$_3$Cl$_2$, Mn$_4$N$_3$Br$_2$, Mn$_4$N$_3$S$_2$, Mn$_4$N$_3$Se$_2$, Cr$_4$N$_3$F$_2$, and Cr$_4$N$_3$O$_2$ exhibit FM ground states with total magnetic moments of 14.6, 14.53, 15.03, 14.6, 12.78, 13.52, 14.7, 8.75, 10.66, 10.6, 8.91, 10.36, 14.87, 16.68, 16.62, 16.61, 14.99, 16.0, 12.89, and 11.1~$\mu_\mathrm{B}$ per unit cell, respectively. The total magnetic moments are consistent with the physical model proposed by Kumar et al.~\cite{Kumar2017}, except for V$_4$C$_3$. For instance, in the case of Mn$_4$C$_3$T$_x$ MXenes, where T$_x$ denotes halogen terminations, the possible oxidation states of Mn are +4 and +3, resulting in 3 and 4 unpaired d-electrons after donating to nearby atoms. This corresponds to the local magnetic moment of 3$\mu_\mathrm{B}$ and 4$\mu_\mathrm{B}$ per Mn atom. Therefore, the expected total magnetic moment of around 14 $\mu_\mathrm{B}$ for Mn groups in halogen-terminated carbides is consistent with our results. However, as the size of the functional group increases, the total magnetic moment also rises, especially in the case of Se, Te, and I. Similarly, the total magnetic moment of nitride MXenes can be explained. The magnetic moment is higher for nitride MXenes than their carbide counterpart due to the presence of three extra electrons per unit cell introduced by nitrogen. It is noteworthy that, apart from Mn$_4$N$_3$S$_2$ and Mn$_4$N$_3$Cl$_2$, none of the other ferromagnetic MXenes identified in this work were predicted in earlier machine-learning studies~\cite{Khatri2023, Xin2023}. Among the remaining materials, we found that Fe and V transition metal MXenes are antiferromagnetic for all functional groups. This indicates that the most preferred magnetic states for MXenes are NM and AFM.

\subsection{Stability analysis}
\begin{figure}[H]
    \centering
    \includegraphics[width=0.5\linewidth]{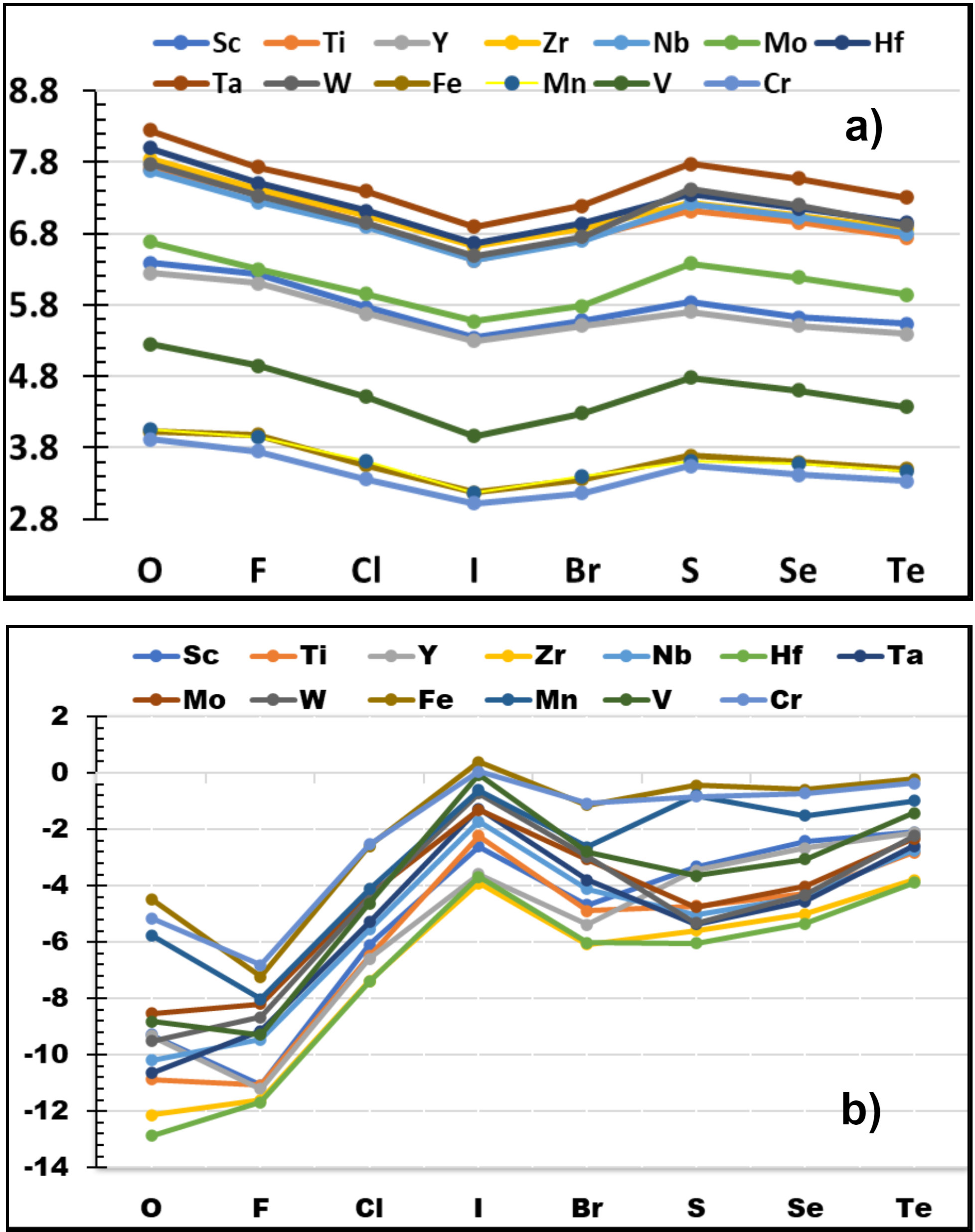}
    \caption{A schematic summary of stability results (a) variation in cohesive energy (eV/atom) and, (b) formation energy (eV/unit cell) with different functional groups for carbide MXenes. (colored lines indicate transition metals)}
    \label{fig:stability}
\end{figure}

To ensure the structural stability of our investigated materials, we calculate cohesive energy, which measures the strength of the forces binding atoms together in a crystal. Additionally, to quantify the chemical bonding strength between transition metal atoms at the surface and terminal groups, we examine their formation energy. The formation (E$_f$) and cohesive ($E_{\mathrm{coh}}$) energy of the materials is calculated using the following equations.

\begin{equation}
E_f = E(\mathrm{M}_4\mathrm{X}_3\mathrm{T}_x) - E(\mathrm{M}_4\mathrm{X}_3) - \frac{2E_{\mathrm{bulk}}(\mathrm{T}_x)}{n},
\end{equation}

\begin{equation}
E_{\mathrm{coh}} = -\frac{1}{m+x+t}\left[ E(\mathrm{M}_4\mathrm{X}_3\mathrm{T}_x) - mE(\mathrm{M}) - xE(\mathrm{X}) - tE(\mathrm{T}_x) \right],
\end{equation}

where $E(\mathrm{M}_4\mathrm{X}_3\mathrm{T}_x)$ and $E(\mathrm{M}_4\mathrm{X}_3)$ denote the total energies of the functionalized and bare MXenes, respectively. $E_{\mathrm{bulk}}(\mathrm{T}_x)$ is the total energy of the functional group in its bulk or gaseous reference state, and $n$ is the number of atoms in that reference unit. $E(\mathrm{M})$, $E(\mathrm{X})$, and $E(\mathrm{T}_x)$ represent the total energies of isolated M, X, and T atoms, respectively, while $m$, $x$, and $t$ correspond to the number of M, X, and T atoms in the unit cell.
The calculated formation and cohesive energy variation with different functional groups for carbide MXenes is illustrated in Figure.~\ref{fig:stability}(a,b). Both energy results indicate that I-terminated MXenes are the least stable among all functional groups. The positive formation energies of Fe$_4$C$_3$I$_2$, Fe$_4$N$_3$I$_2$, Cr$_4$C$_3$I$_2$, V$_4$C$_3$I$_2$, and Fe$_4$N$_3$Te$_2$ further intensify the instability of this group. The lattice constant, monolayer thickness and M-Tx bond length are found to increase with the increase in the atomic radii of functional groups. The M-Tx bond length is larger for the I terminal group and smallest for the O and F groups. Larger bond lengths typically reduce structural stability due to weaker overlap between atomic orbitals. This is why I terminated MXenes tend to be the least stable. 
\\
Our formation and cohesive energy results for Ti$_4$C$_3$O$_2$, Ti$_4$C$_3$F$_2$, Ti$_4$N$_3$O$_2$, and Ti$_4$N$_3$F$_2$ are in good agreement with previous works \cite{Huang2023}, \cite{Zhang2018}, \cite{Urbankowski2016}. Cohesive energy for Nb, Hf, Ta, Zr, W, and Ti range from 8.25 to 6.5 eV/atom, while those for Y, Sc, and Mo fall between 6.7 and 5.3 eV/atom. These values indicate that these compounds are stable across all functional groups. We also noticed that cohesive energy is slightly higher for carbides than nitride counterparts. It can be seen in Figure.~\ref{fig:stability}(a,b) that Cr, Mn and Fe transition metal MXenes have lower formation and cohesive energy than others. Therefore, to further confirm the structural stability and investigate the possibility of synthesizing the studied FM MXenes, we evaluate their dynamical stability. In this study, we employed the finite displacement method to calculate phonon dispersion using PHONOPY software \cite{Togo2023}. All FM MXenes are dynamically stable except Mn$_4$C$_3$I$_2$, Cr$_4$C$_3$F$_2$, Cr$_4$C$_3$Cl$_2$, and Cr$_4$N$_3$O$_2$. Among the 20 FM MXenes, 13 were Mn-based and 7 were Cr-based. Of the Cr-based, 3 are found to be unstable. This suggests that Mn-based M$_4$X$_3$T$_x$ MXenes are more stable and preserve FM states irrespective of the functional group.

\subsection{Electronic Properties}
\begin{figure}[H]
        \centering
        \includegraphics[width=1\linewidth]{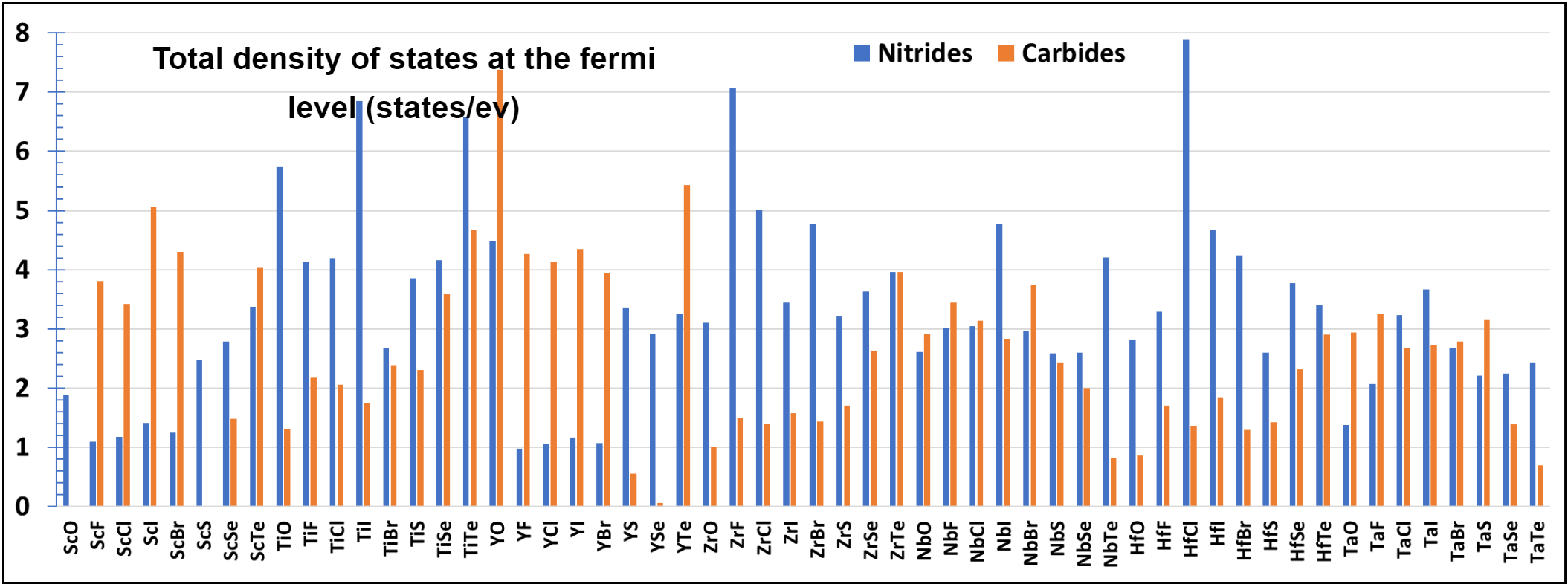}
        \caption{Comparison of the density of states at the Fermi level for carbide and nitride MXenes (blue bars are nitrides and yellow bars indicate carbides, on the x-axis compositions are in the form M$_4$T$_x$ i.e, first letter is transition metal and the second indicates functionalized group}
        \label{fig:dos}
\end{figure}

Moving further to understand the electronic characteristics of these stable materials, we performed density of states (DOS) calculations for their respective ground state. Due to the presence of unpaired 3/4/5d electrons, most MXenes are metallic in different layer categories. In this work, we found that all bare and Sc, Ti, Y, Zr, Nb, Hf, Ta, Fe, V, Mo and W transition metal M$_4$X$_3$T$_x$ MXenes exhibit metallic behavior, except for Sc$_4$C$_3$O$_2$, Sc$_4$C$_3$S$_2$, Y$_4$C$_3$S$_2$, and Y$_4$C$_3$Se$_2$. The total density of states at the Fermi level ($N(E_F)$) for carbide and nitrides is illustrated in Figure.~\ref{fig:dos}. It can be seen that nitrides have high $N(E_F)$ compared to carbides, which is correlated to the one extra electron on each N atom. The difference between nitride and carbide $N(E_F)$ is higher for Ti, Hf and Zr and lower for Nb and Ta. In these materials, d-orbitals of transition metals are the main contributors. Due to their excellent electronic conductivity and large interlayer spacing, Ti, Ta and Nb-based M$_4$X$_3$O$_2$/F$_2$ MXenes have shown improved performance for Li-ion batteries \cite{Hussain2024}.

\begin{figure}[H]
    \centering
    \includegraphics[width=1\linewidth]{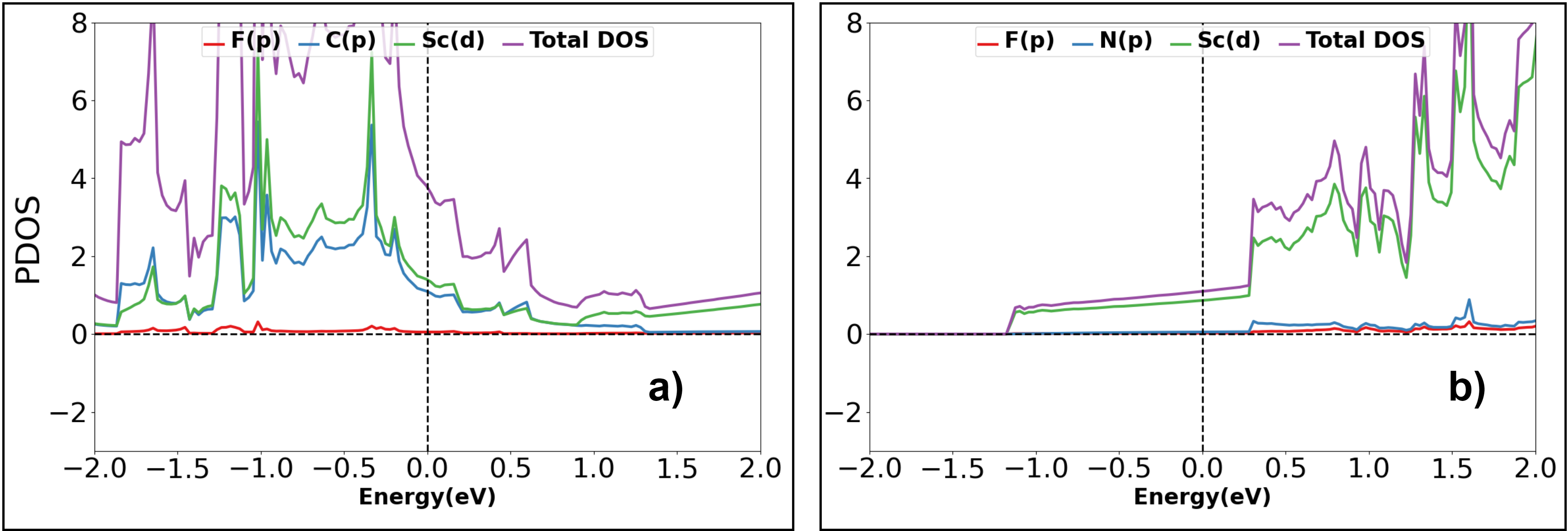}
    \caption{PDOS of carbide and nitride Sc-based MXenes to explain their reverse metallic behavior (a) Sc$_4$C$_3$F$_2$ and (b) Sc$_4$N$_3$F$_2$}
    \label{fig:5}
\end{figure}

However, we further observe that F-, Cl-, I-, and Br-terminated Y- and Sc-based MXenes exhibit a reverse trend for $N(E_F)$, wherein carbide MXenes possess significantly larger $N(E_F)$ than their nitride counterparts (see Figure.~\ref{fig:dos}). This behavior can be understood using arguments similar to those discussed in the magnetism section. Considering the electronegativity sequence F $>$ O $>$ Cl $>$ N $>$ Br $>$ I $>$ S $\approx$ Se $\approx$ C $>$ Te, substantial charge redistribution occurs upon surface functionalization. Upon halogen termination, for the Sc$_4$C$_3$ systems (as there are no unpaired electrons left after sharing to C), the higher electronegativity of F, Cl, Br, and I relative to C results in enhanced attraction of Sc $d$ electrons toward the surface functional groups, leading to strong hybridization between Sc-3$d$ and C-2$p$ orbitals. This is evident from the projected density of states shown in Figure.~\ref{fig:5}(a), where comparable contributions from Sc-3$d$ and C-2$p$ orbitals appear at the Fermi level. Similar PDOS features are observed for all halogen-terminated Sc-based carbides. In contrast, for nitride MXenes, the remaining Sc $d$ electrons are primarily shared with the surface functional groups, resulting in reduced hybridization with N $p$ states near $E_F$, as shown in Figure.~\ref{fig:5}(b). Based on this trend, Sc$_4$C$_3$O$_2$, Sc$_4$C$_3$S$_2$, Y$_4$C$_3$S$_2$, and Y$_4$C$_3$Se$_2$ might be expected to be metallic; however, they instead exhibit semiconducting behavior with narrow band gaps of 0.02~eV, 0.133~eV, 0.31~eV, and 0.23~eV, respectively. 

Focusing on the stable FM MXenes, we find that Mn$_4$N$_3$S$_2$, Mn$_4$N$_3$O$_2$, Mn$_4$N$_3$F$_2$, Mn$_4$N$_3$Cl$_2$, Mn$_4$N$_3$Br$_2$, Cr$_4$N$_3$F$_2$, Mn$_4$C$_3$Cl$_2$, Mn$_4$C$_3$Br$_2$, and Mn$_4$C$_3$S$_2$ exhibit band gaps of 0.7, 2.0, 2.5, 2.0, 1.4, 2.4, 0.8, 0.6, and 0.25~eV, respectively, in the minority spin channel. The coexistence of metallic transport in the spin-up channel and a band gap in the minority spin channel results in 100\% spin polarization, classifying these materials as half-metals, which are highly desirable for spintronic applications.  In other FM MXenes, the minority spin channel is metallic; however, the density of states at the Fermi level, $N(E_F)$, is larger for the spin-up channel compared to the minority spin channel. The spin polarization, $P$, in these materials is calculated as:  
\begin{equation}
P = \frac{N(E_F)^\uparrow - N(E_F)^\downarrow}{N(E_F)^\uparrow + N(E_F)^\downarrow},
\end{equation}
where $N(E_F)^\uparrow$ and $N(E_F)^\downarrow$ represent the density of states at the Fermi level for the majority and minority spins, respectively.  
We find that Cr$_4$C$_3$O$_2$, Cr$_4$C$_3$S$_2$, Cr$_4$C$_3$Se$_2$, Mn$_4$C$_3$F$_2$, Mn$_4$C$_3$Se$_2$, Mn$_4$C$_3$Te$_2$, and Mn$_4$N$_3$Se$_2$ possess spin polarizations of 70\%, 75\%, 26\%, 50\%, 56\%, 58\%, and 62\%, respectively. The thickness of these FM materials ranges from 10 to 12~\AA, while their magnetic moments per formula unit fall between 14 and 16~$\mu_\mathrm{B}$. These combined electronic and magnetic properties make these materials promising candidates for low-dimensional spintronic devices.

\section{Conclusion}

In conclusion, we employ a machine-learning-assisted high-throughput DFT framework that provides an efficient and scalable route for exploring MXenes at large scale. By incorporating ML-predicted lattice parameters as initial inputs to first-principles calculations, the computational cost associated with structural convergence is substantially reduced, enabling systematic screening of complex higher-order MXenes. Using this approach, we identify 16 Cr- and Mn-based MXenes as stable ferromagnets with magnetic moments of 14--16~$\mu_\mathrm{B}$ per unit cell and thicknesses of 10--12~\AA, highlighting their suitability for low-dimensional spintronic applications. In addition, Y$_4$C$_3$O$_2$, Y$_4$N$_3$O$_2$, and Y$_4$N$_3$S$_2$ exhibit weak ferromagnetism that may be further engineered via a double transition-metal strategy. While most MXenes are metallic, Sc$_4$C$_3$O$_2$, Sc$_4$C$_3$S$_2$, Y$_4$C$_3$S$_2$, and Y$_4$C$_3$Se$_2$ are identified as narrow-band-gap semiconductors. Overall, this systematic investigation of 234 MXenes establishes both a comprehensive dataset and a scalable ML-assisted DFT framework, providing valuable insights into composition–property relationships and a robust foundation for future large-scale discovery of MXene-based spintronic and functional materials.

\section{Conflicts of interest}
The authors declare that they have no conflict of interest
\section{Acknowledgment}
The author would like to thank Prof. Arti Kashyap for providing research facilities and research guidance. I also thank the National PARAM super-computing facility at C-DAC, India and IIT Mandi.
\section{Data Availability}
The data supporting the findings of this study are available within the article. Additional data related to this work are available from the corresponding author upon request.

{\small
\bibliographystyle{IEEEtran}
\bibliography{refernces}
}

\end{document}